\documentclass[sn-mathphys,Numbered,11pt]{sn-jnl}
\usepackage{graphicx}%
\usepackage{multirow}%
\usepackage{amsmath,amssymb,amsfonts}%
\usepackage{amsthm}%
\usepackage{upgreek}%
\usepackage{subfigure}
\usepackage{mathrsfs}%
\usepackage[title]{appendix}%
\usepackage{xcolor}%
\usepackage{textcomp}%
\usepackage{manyfoot}%
\usepackage{booktabs}%
\usepackage{algorithm}%
\usepackage{algorithmicx}%
\usepackage{algpseudocode}%
\usepackage{listings}%
\theoremstyle{thmstyleone}%
\theoremstyle{thmstyletwo}%

\theoremstyle{thmstylethree}%

\begin{document}

\title[Developments on frequency domain multiplexing readout for large arrays of transition-edge sensor X-ray micro-calorimeters]{Developments on frequency domain multiplexing readout for large arrays of transition-edge sensor X-ray micro-calorimeters}

\author[1*]{D.~Vaccaro}
\author[1,2]{H.~Akamatsu}
\author[1]{L.~Gottardi}
\author[1]{M.~de~Wit}
\author[1]{M.P.~Bruijn}
\author[3]{J.~van~der~Kuur}
\author[1]{K.~Nagayoshi}
\author[1]{E.~Taralli}
\author[1]{K.~Ravensberg}
\author[1,4]{J-R.~Gao}
\author[1,5]{J.W.A.~den~Herder}

\affil[1]{NWO-I/SRON Netherlands Institute for Space Research, 2333 CA Leiden, The Netherlands}
\affil[2]{QUP International Center for Quantum-field Measurement Systems for Studies of the Universe and Particles, Fuji G07, 1-1 Oho, Tsukuba, Japan}
\affil[3]{NWO-I/SRON Netherlands Institute for Space Research, 9747 AD Groningen, The Netherlands}
\affil[4]{Optics Group, Delft University of Technology, Delft, 2628 CJ, The Netherlands}
\affil[5]{Universiteit van Amsterdam, Science Park 904, 1090GE Amsterdam, The Netherlands}

\abstract{At SRON we have been developing X-ray TES micro-calorimeters as backup technology for the X-ray Integral Field Unit (X-IFU) of the Athena mission, demonstrating excellent resolving powers both under DC and AC bias. We also developed a frequency-domain multiplexing (FDM) readout technology, where each TES is coupled to a superconducting band-pass LC resonator and AC biased at MHz frequencies through a common readout line. The TES signals are summed at the input of a superconducting quantum interference device (SQUID), which performs a first amplification at cryogenic stage. Custom analog front-end electronics and digital boards take care of further amplifying the signals at room temperature and of the modulation/demodulation of the TES signals and bias carrier, respectively.

We report on the most recent developments on our FDM technology,  which involves a two-channel demonstration with a total of 70 pixels with a summed energy resolution of 2.34$\pm$0.02~eV at 5.9~keV without spectral performance degradation with respect to single-channel operation. Moreover, we discuss prospects towards the scaling-up to a larger multiplexing factor up to 78 pixels per channel in a 1-6~MHz readout bandwidth.}


\maketitle

\section{Introduction}\label{intro}

The last decades saw a growing interest in cryogenic detectors for many physics fields, from astro-particle physics and cosmology to plasma and material science, in particular regarding transition-edge sensors (TES) \cite{tes}, due to their attractive characteristics of non-dispersive single-photon counting capabilities, high spectral resolution and quantum efficiency, large photon collecting area by means of arrays with large number of pixels.

A TES is a superconducting film strongly coupled to a radiation absorber and weakly linked to a thermal bath at a temperature $T_b < T_{\text{C}}$, where $T_{\text{C}}$ is the critical temperature of the film, typically at a level of 100~mK. TESs are used as extremely sensitive thermometers, by exploiting the sharp phase transition of the superconducting film. Under stiff voltage bias (either ac or dc), the TES is heated to its $T_{\text{C}}$ and stable operation is granted by the so-called negative electro-thermal feedback (ETF) mechanism. The observable signal is the TES current, which is typically amplified at cryogenic stage by a Superconducting QUantum Interference Device (SQUID). 

At SRON we have been developing X-ray TES micro-calorimeters as backup technology for Athena X-IFU,  optimizing in the course of the years the fabrication process \cite{ken} and design to achieve state-of-the-art energy resolution for both ac- and dc-readout \cite{weaklink,mdw2022}, demonstrating levels of crosstalk \cite{tct} and environmental sensitivities \cite{gsens,sgp} compliant with requirements for satellite astronomical missions.

\begin{figure}[!h]
\centering
\includegraphics[angle=90,origin=c,width=0.55\textwidth]{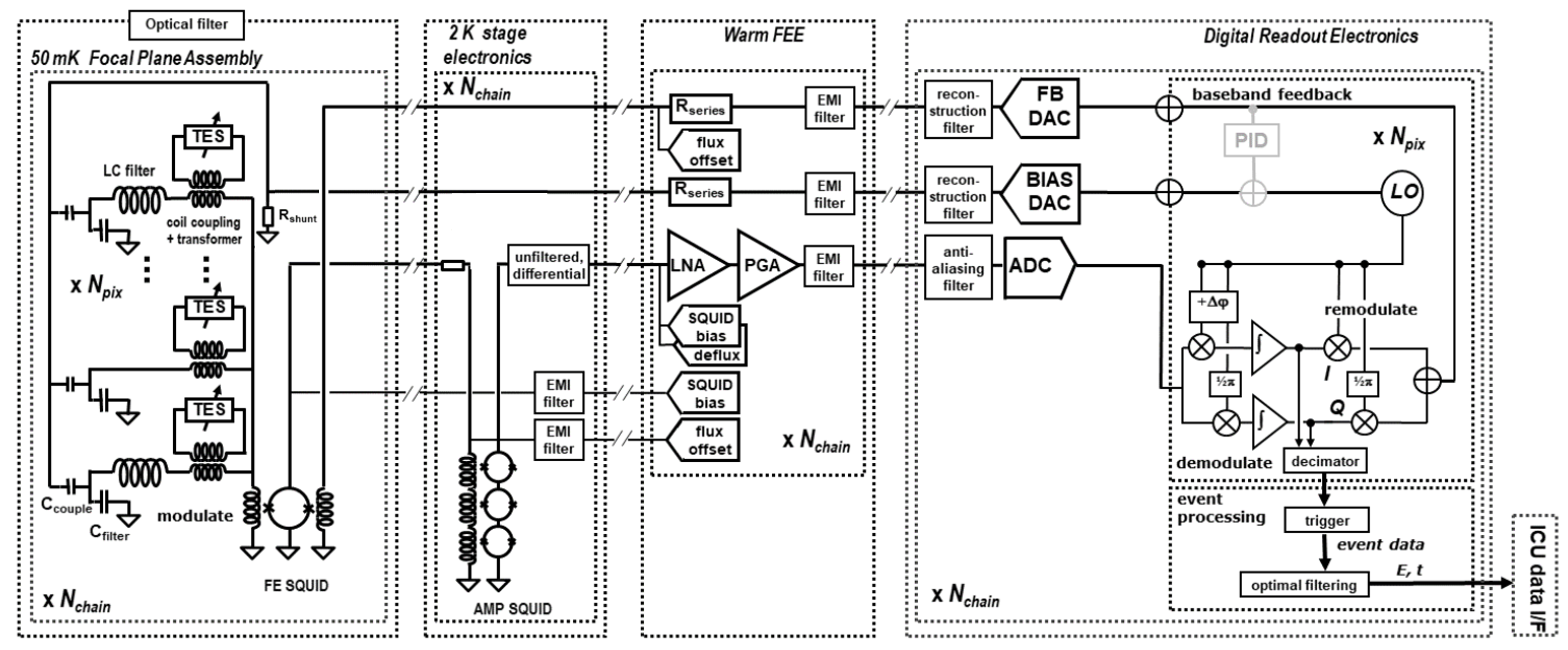}
\caption{Schematic representation of the FDM readout with base-band feedback, with a highlight on the electronics used at the different temperature stages. Adapted from \cite{dac} with authorization from the authors.}\label{bbfb}
\end{figure}

We have also been developing a frequency domain multiplexing (FDM) technology to readout TES large arrays with a base-band feedback scheme \cite{bbfb} (see Figure~\ref{bbfb}). A multiplexing readout scheme is necessary for experiments employing hundreds or thousands of detectors, to minimize harness complexity and the thermal load at cryogenic stage, particularly for space missions where these requirements are strict. In the FDM scheme, the detectors are voltage biased with a comb of sinusoidal carriers at different frequencies in the MHz range. Each TES is put in series with a tuned, high-$Q$ LC band-pass filter to limit the information bandwidth and allow only one carrier from the comb to provide the bias. To keep the same electrical bandwidth across the resonators, the inductance is kept constant and the frequencies are tuned by changing the capacitance. The TESs signals are then summed at the input coil of a SQUID and demodulated by room temperature electronics. The base-band feedback nulls the signal at the SQUID by feeding back to the SQUID feedback coil the sum of the TES signals after a phase compensation.

In this contribution we report on the most recent developments on our FDM readout technology, which consist in a dual-channel system performance demonstration.

\section{Experimental setup}

\begin{figure}[!h]
\centering
\includegraphics[width=0.9\textwidth]{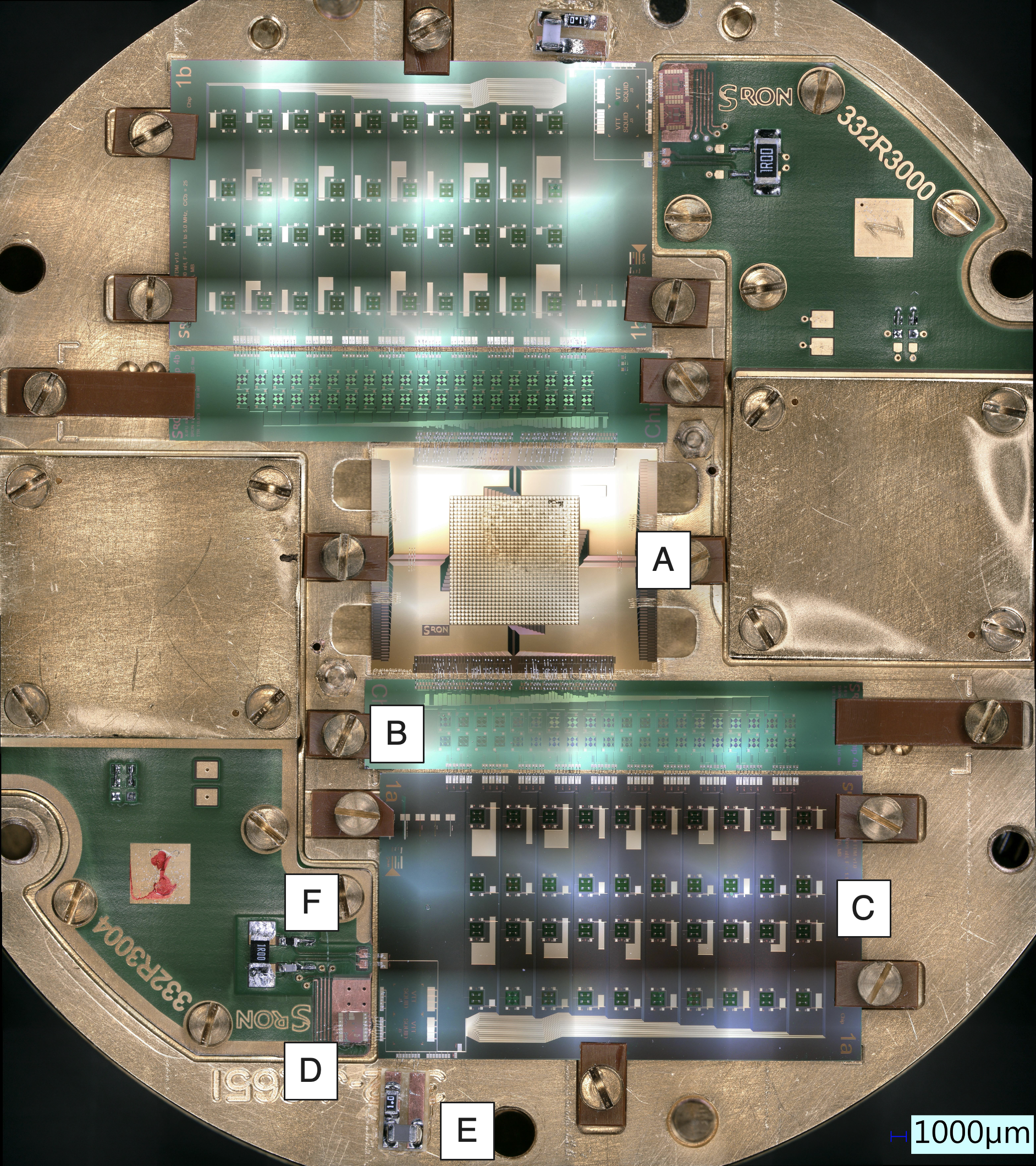}
\caption{Top view of the 50~mK assembly of the setup holding the cryogenic components, with the two populated channels. Visible are the TES array [A], transformer [B] and LC filter [C] chips, the first-stage SQUID [D] with the RC low-pass filter[E] and the bias SMD resistor [F].}\label{40pxA}
\end{figure}

The cryogenic part of the setup is depicted in Figure~\ref{40pxA}.  The setup hosts a $32\times32$ TES array. Each TES consists of a Ti/Au bilayer of $80 \times 13~\upmu$m~dimensions in the north side and $80 \times 10~\upmu$m~dimensions in the south side, deposited on a 500~nm thick membrane. The bilayer is coupled to a $240 \times 240~\upmu$m, $2.3~\upmu$m thick gold absorber ($C = 0.85$~pJ/K at 90~mK). The TES have sheet resistance $R_{\square} \approx 25$~m$\Omega$, $T_C \approx 83$~mK and $G \approx$~60~pW/K. 

For FDM readout, the TES chip is connected to custom LC filter and transformer chips.  Both channels of the setup are populated with FDM cryogenics components, $i.e.$ LC filter chips, transformer chips front-end (FE) and amplifier (AMP) SQUIDs. For both channels, the inductance of the LC filter chips is $2 \upmu$H and the transformer coil ratio is 1:0.875 with coupling factor $k = 0.94$, making for an effective inductance at the TES of $3 \upmu$H.  The use of transformers with different coil ratio allows to fine tune the readout circuit parameters, matching the TES impedance at setpoint, without having to fabricate different LC filter chips, which is a complicated lithographic process, while transformer chips require simpler fabrication steps. 40 different LC resonances are obtained in the range between 1~MHz and 5 MHz by varying the capacitance in the range from 0.5~nF to 10~nF. The detectors are voltage-biased via a 750~m$\Omega$ resistor and a capacitive divider with 1:25 ratio, resulting in an effective shunt resistance of $\approx 1$~m$\Omega$.

The signals from the TES are summed at the input coil of a SQUID amplifier system. We are using in both channels a two-stage SQUID system developed by VTT \cite{vtt}, consisting of a J3 model as first-stage and a L1x model as second-stage. Typical power dissipation for the SQUID models used is $\simeq 0.3$~nW for the first stage SQUID and $\simeq 300$~nW for the second stage SQUID. Typical readout current noise referred at SQUID input is about 4~pA/$\sqrt\text{Hz}$. A RC low-pass filter is implemented at the input coil of the first-stage SQUID to suppress electrical instabilities due to the coupling between the SQUID input inductance and parasitics in the LC filter chip.

The cryogenic components are mounted on an oxygen-free high-conductivity (OFHC) Cu holder and enclosed in an aluminium casing. The holder is hosted in a Leiden Cryogenics (model CF-CS51-400) dilution refrigerator unit with $400 \upmu$W cooling power at 120 mK. The setup is suspended via Kevlar wires from the mixing chamber to dampen mechanical oscillations \cite{kevlar} and thermalized to it via OFHC Cu braids.  The temperature of the setup is monitored via a Ge thermistor and regulated via a 500 $\Omega$ heater through a software PID loop.  The tests reported in this paper were performed at a base temperature of 50~mK, with a stability at a level of $1 \upmu$K.

A custom digital "DEMUX" board mounting a Xilinx Virtex 7 FPGA handles the generation of the TES bias voltages modulated by the ac carriers, the demodulation of the SQUID output signal and re-modulation for base-band feedback.  The SQUID control and low-noise amplification of the SQUID output is performed by a custom front-end electronics (FEE).

To test the detectors, the setup is coupled to X-ray sources. For spectral performance characterization typically we use $^{55}$Fe sources mounted on the cold aluminium casing to directly illuminate the TES with X-rays of the Mn-K$\upalpha$ spectral complex ($E \sim 5.9$ keV).  The count rate is tuned via layers of aluminium foil placed in front of the source. For the measurements reported here, the count rate is about 1 count per second per pixel.

Apart from cryogenic component availability, the expansion from one to two channels did not require any specific further effort: our FEEs are already predisposed for a two-channel configuration and the software controlling the system settings is designed for scalability and automation.  In order to set the system for measurements, the steps required are: 1) Network Analyzer scan to identify the LC resonances, 2) calibration of the base-band feedback phase delay, 3) TES IV curve measurement and calibration,  4) Frequency Shift Algorithm (FSA) \cite{FSA} calibration to mitigate the impact of intermodulation distortions and 5) lock of the feedback with pixels on their quiescence point.  The full process takes few hours, mainly due to the FSA calibration.  All steps are fully automated and the first four are only required when new components are installed in the system.  Setting up the two-channel system once the settings (bias voltages, bias frequencies, phase delays, FSA) are known requires only few minutes.

\section{Results}

Demonstrations of single-channel FDM readout of TES micro-calorimeters have been reported in \cite{hirokidemo}.  Towards a two-channel demonstration, cryogenic components were selected to populate the second readout channel to be as consistent as possible with the first channel. Due to component availability, we could not use LC filter and transformer chips with 100\% yield. From the total of 40 pixels available per channel, 6 and 4 from the first and second channel respectively were not included in the experiment, due to known lithographic defects in either the transformer chip or LC filter chip. This is the only limiting factor to the multiplexing factor, preventing us to measure the nominal total of 80 pixels across the two readout channels.

To verify the system performance in dual-channel operation, we first characterized each channel individually, biasing the TES at about 15\% of their superconducting transition while keeping the TES and SQUIDs in the other channel in the superconducting state. 

\begin{figure}[h]%
\centering
\subfigure[]{\includegraphics[width=0.49\textwidth]{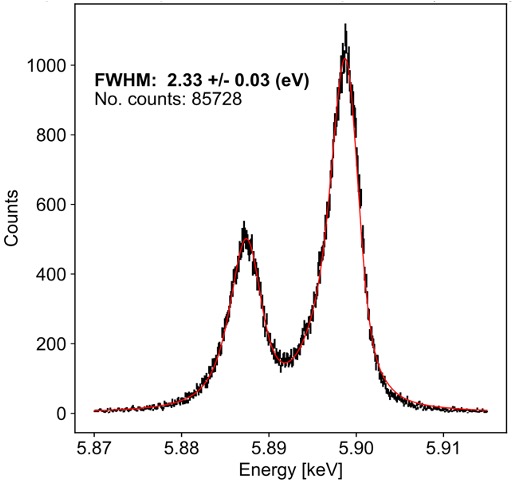}}
\subfigure[]{\includegraphics[width=0.49\textwidth]{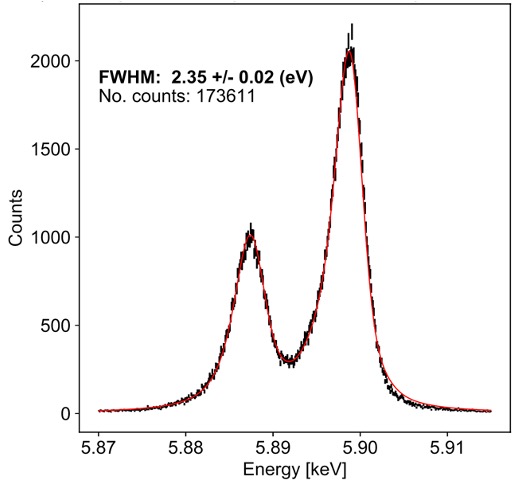}}
\caption{Energy spectra from the $^{55}$Fe source collected with the simultaneous operation of the two-channels of the setup, the first with 34-pixel multiplexing (a) and the second with 36-pixel multiplexing (b).  Data are represented in black and the best fit is represented in red.}\label{2chdemo}
\end{figure}

The experiment consisted in collecting several thousands of X-rays per pixel from a $^{55}$Fe source mounted on the Al casing of the setup.  Long term temperature drifts are corrected by using the TES baseline current and the pulse height information.  For each pixel, the energy of the X-ray events is estimated using the optimal filtering technique. Then, the ``summed" spectral performance of all the pixels is calculated by fitting the Mn-K$\upalpha$ model \cite{holzer} to the histogram of the collected events by minimizing the Cash statistics in the maximum-likelihood method \cite{cstat}.  The non-linearity of the gain scale is also corrected, by using the zero energy, MnK$\upalpha$ and MnK$\upbeta$ information. 

In this fashion,  we measured energy resolutions of $2.29\pm0.03$ eV (34-pixel multiplexing) and $2.31\pm0.03$~eV (36-pixel multiplexing) for the first and second channel, respectively. The average single-pixel performance for the TES has been measured to be at a level of 2~eV. The $\sim$~1~eV degradation in quadrature is consistent with what observed in the single-channel demonstrations, and we attribute it mainly to residual effects of AC Josephson effects at high bias frequencies and intermodulation distortions. Both are not fundamental limitations, as they can be solved with higher $R_N$ TES and optimization of the readout circuit. For this we need the fabrication of new components, which is currently under way.

With this reference, we then characterized the performance of the system with both channels and all 70 pixels active, to verify their simultaneous operation would not generate any interference spoiling the energy resolution. The results are reported in Figure~\ref{2chdemo}. Overall,we measured a summed performance of $2.34\pm0.02$ eV with 2 FDM channels and 70 pixels in total. We observed no degradation with respect to single-channel measurements, demonstrating for the first time the multi-channel operation of our FDM technology.

\section{Summary and future outlook}

We reported on recent developments at SRON on the frequency domain multiplexing readout for TES X-ray microcalorimeters for future space-borne astronomical observatories and ground-based applications, such as plasma and material science.

Starting from our previous one-channel demonstrations, we populated with cryogenic components both channels of an FDM demonstrator setup. We measured a summed spectral performance of $2.34\pm0.02$ eV at 5.9~keV for a total of 70 pixels with no degradation from single-channel operation.  The main limiting factor regarding multiplexing number and performance is attributed to the components (TES array, LC filters, transformers) currently being available.

Activities are being carried out to further improve the multiplexing capabilities of our FDM readout, by fabricating LC filter chips to allow for multiplexing factors up to 78 and to extend the readout bandwidth up to 6 MHz. The main drawbacks of this configuration would be electrical crosstalk (carrier leakage) and higher weak-link effect, which can both be mitigated by using TES with higher resistance.  We expect to fabricate the new batch of LC filters in mid-2024, so that first experiments with 78-pixel FDM readout per channel should occur before the end of next year.

\section*{Acknowledgements}
SRON is financially supported by the Nederlandse Organisatie voor Wetenschappelijk Onderzoek.

This work is part of the research programme Athena with project number 184.034.002, which is (partially) financed by the Dutch Research Council (NWO).

The SRON TES arrays used for the measurements reported in this paper is developed in the framework of the ESA/CTP grant ITT AO/1-7947/14/NL/BW.

\section*{Data availability}

The corresponding author makes available the data presented in this paper upon reasonable request.

\end{document}